# Parallelisation of algorithms for *ab initio* computation of material properties

G.-M. Rignanese, J.-M. Beuken, J.-P. Michenaud, and X. Gonze.

*Unité de Physico-Chimie et de Physique des Matériaux, Université Catholique de Louvain,*

*1 Place Croix du Sud, B-1348 Louvain-la-Neuve, Belgium*

(November 1, 1995)

## Abstract

Parallel algorithms for *ab initio* calculations of vibrations modes of solids are presented and implemented under PVM. Load balancing and communication problems are dealt with in order to increase parallelism efficiency. For accurate time measurements, synchronisation of processes must be achieved. The results obtained by working with 1,2,4 and 8 processors of a Convex MetaSeries computer are presented, showing a very good efficiency. Further parallelisation of the codes (involving a three-dimensional Fast Fourier Transform) with compiler directives on a Convex Exemplar, is discussed.

## INTRODUCTION

Recently, theoretical and algorithmic advances have made it possible to determine thermodynamical properties of solids (such as specific heat or thermal-expansion coefficient) from first principles calculations (based on Electromagnetism and Quantum Mechanics) [1,2]. As the thermodynamic functions of a solid are mostly determined by the vibration spectrum of the lattice, the quantities known as interatomic force constants [IFC's], describing at linear order the force created on one nucleus by the displacement of another from its equilibrium position, must be calculated.

This problem is very well suited for a coarse-grain parallelisation. Indeed, the calculation of the IFC's necessitates to treat the perturbation due to the displacement of each atom of the unit cell in each direction of the space. As these perturbations are independent, they can be treated independently by different processors working in parallel.

Typically, after an initialisation step that could last from a few minutes to one hour on a HP-735 or IBM RS6000-550 workstation, a set of six to about one hundred responses will be computed, each of which could need from one hour to several days of CPU time. When all the response calculations are completed, a final step (that could last from a few minutes to a few hours) will gather the results and perform some further analysis. There are no communications between the different response calculations, but a large amount of input data must be made available to them in the initialisation process (mostly the same data for each response). Also a large quantity of results from each response calculation must be treated in the final step.

In the present paper, we first explore the above-mentioned coarse-grain parallelisation, using a master and slaves algorithm. We achieve an excellent speedup on a Convex MetaSeries computer, with up to 8 processors. This is due to the very low sequential fraction of the code and the low amount of communications compared with the computation time. Nevertheless, some limitations of this parallelisation are already apparent: first, the number of such responses is not scalable, and can be small even for some heavily demanding problems; second, the time needed for each response computation may vary (although not by much: 10% to 20%). The first limitation is the most embarassing, since one will not be able to take advantage of the power delivered by e.g. 8 processors or more, in the case of a computation of 7 responses or less. Supposing now that the number of responses is a multiple of the number of processors, which suppress this first problem, the second limitation then becomes apparent, since some processors will stay idled at the end of their work, because their workload is smaller than that of some other processors. This parallelisation is thus very efficient in some cases, but too coarse for a large number of processors.

This is the reason why we also explore another parallelisation of the code, at a much finer level. From an analysis of the algorithm used for the response calculation, we can see that a three-dimensional Fast Fourier Transform [FFT] (for sizes that are typically comprised between $24*24*24$ and $120*120*120$) is the crucial step of this parallelisation. We briefly describes the parallelisation algorithm, its implementation on an Convex Exemplar, and perform timing tests with up to 4 dedicated processors. The cache behaviour is found to be crucial, and an interesting speedup can be obtained for the larger size FFT's, although not as good as for the coarse-grain parallelisation. Interestingly both coarse-grain and fine-grain parallelisations could be combined, thus alleviating partially the disadvantages of each of them.

## I. COARSE-GRAIN PARALLELISATION

In the sequential RESPFN (response function) code, the LOPOL3 procedure deals with the calculation of IFC's. It consists of a single process (task) executed on a single processor (host), that treats all the perturbations due to the displacement of each atom of the unit cell in each direction of the space.

The main part of LOPOL3 routine is composed of two nested loops: the first iterates on the atoms of the unit cell ($i_{atom} = 1 \longrightarrow n_{atom}$), and the second on the directions of the space ($i_{dir} = 1 \longrightarrow 3$). The inner loop consists of the TREAT routine which treats a given perturbation ($i_{atom}, i_{dir}$) if necessary. This algorithm is reproduced in Fig. 1.

```
procedure LOPOL3 (in data, n_atom)
begin
    do i_atom = 1 ⟶ n_atom
    /* Loop on the atoms */
        do i_dir = 1 ⟶ 3
        /* Loop on the directions */
            if TEST(i_atom, i_dir) then
            /* Tests if a given perturbation should be treated */
                TREAT(data, i_atom, i_dir)
                /* Treats one perturbation */
            fi
        od
    od
end
```

FIG. 1. Sequential algorithm of the main part of the LOPOL3 routine.

In this first section, we want to present the parallelisation process that has lead us to the PARESPFN (parallel response function) code consisting of $n_{task}$ different processes executed on $n_{host}$ processors. In this parallel algorithm, each process treats the perturbation due to the displacement of one atom of the unit cell.

To build the parallel code, we adopt a "master & slaves" structure, where the master distributes the perturbations between the slaves, and each slave treats one or more perturbations. This struture is particularly well suited for successive approaches.

First of all, we associate a process (which) to each given perturbation, defined by ($i_{atom}, i_{dir}$), that has to be treated. This is done by the CORRESPOND routine (reproduced in Fig. 2), which creates two arrays pratom and prdir containing respectively $i_{atom}$ and $i_{dir}$ for each process.

Then, there are many ways to distribute these processes between the slaves. Let us analyze two of them paying special attention to load balancing and communication problems.

The first approach, the most simple, consists of distributing one process by slave. In this case, there exists a univocal correspondence between a process (which) and the slave (who) that is in charge of it, as suggested in Fig. 3. If the slave knows who he is (me), he knows which task he must execute.

```
procedure CORRESPOND (in n_atom, out n_task, pratom, prdir)
begin
    n_task = 0
    do i_atom = 1 ⟶ n_atom
    /* Loop on the atoms */
        do i_dir = 1 ⟶ 3
        /* Loop on the directions */
            if TEST(i_atom, i_dir) then
            /* Tests if a given perturbation should be treated */
                n_task = n_task + 1
                pratom(n_task)=i_atom
                prdir(n_task)=i_dir
            fi
        od
    od
end
```

FIG. 2. Algorithm the CORRESPOND routine which creates two arrays pratom and prdir containing respectively $i_{atom}$ and $i_{dir}$ for each process.

The algorithms of LOPOL3 procedure (master) and of the program SLAVE_OF_LOPOL3 are reproduced in Figs. 5 and 6, using the generic language described in Fig. 4 for master/slaves communication.

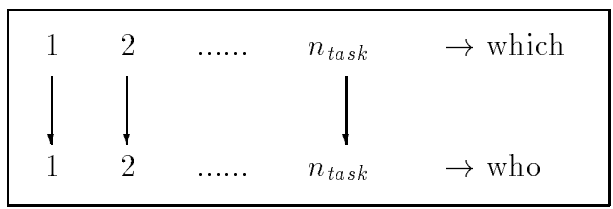

FIG. 3. Distribution of the processes between the slaves in the case of a distribution "one process by slave".

This method leads to two important inconveniences, when the number of processes $n_{task}$ is bigger than the number of processors $n_{host}$, which is generally the case. First, this leads to an overloading of some processors, which is added to the pre-existing load. Second, as the data must be given to all slaves, these can sent twice to the same processor (on which two or more slaves are running). These are sources of inefficiency.

- **SPAWN (in $n_{slave}$, out slave)** :
  create $n_{slave}$ slaves identified by the array slave($1 \rightarrow n_{slave}$).

- **SEND (in who, what)** :
  send what to who, where what is a set of data, and who is whether the master or one or more slaves.

- **RECEIVE (in who, what)** :
  receive what from who, where what is a set of data, and who is whether the master or one or more slaves.

- **ME (in slave, $n_{slave}$, out me)** :
  determine between the $n_{slave}$ slaves, identified by the array slave, who I am (me).

- **CONFIG (out $n_{host}$)** :
  determine the total number of available processors ($n_{host}$).

FIG. 4. Generic language for master/slaves communications.

```
procedure LOPOL3_MASTER (in data, n_atom)
begin
   CORRESPOND (n_atom, n_task, pratom, prdir)
   SPAWN (slave, n_task)
   SEND (slave(1 ⟶ n_task), {n_task, slave, data, pratom, prdir})
   /* Send data and various informations to all slaves*/
   do i_task = 1 ⟶ n_task
   /* Loop on the slaves */
      RECEIVE (slave(1 ⟶ n_task), {who})
      /* Wait until a slave has finished */
      SEND (slave(who), {who})
      /* Send an ending message to slave(who) */
   od
end
```

FIG. 5. Algorithm the LOPOL3 routine in the case of a distribution "one process by slave".

```
program LOPOL3_SLAVE
begin
   RECEIVE (master, {n_task, slave, data, pratom, prdir})
   /* Receive data and various informations from master */
   ME (slave, n_task, me)
   /* Determine who I am */
   i_atom=pratom(me)
   i_dir=prdir(me)
   /* Determine which perturbation I must treat */
   TREAT(data, i_atom, i_dir)
   SEND (master, {me})
   /* Send a message to the master to tell him I have finished */
   RECEIVE (master, {me})
   /* Wait an ending message from master */
end
```

FIG. 6. Algorithm the program SLAVE_OF_LOPOL3 in the case of a distribution "one process by slave".

We now develop a second approach in order to avoid these problems and to improve the efficiency. This method consists of distinguishing process (which) and slave (who) by limiting the number of slaves $n_{slave}$ to the number of processors $n_{host}$ available.

At the beginning, the master distributes one process by slave. Then when a slave has finished, he asks the master for a new task. This distribution "at request" is more complex. Indeed, in this case, there does not exist anymore a univocal correspondence between a process (which) and the slave (who) that is in charge of it, as suggested in Fig. 7. So the slave must be told which process to execute.

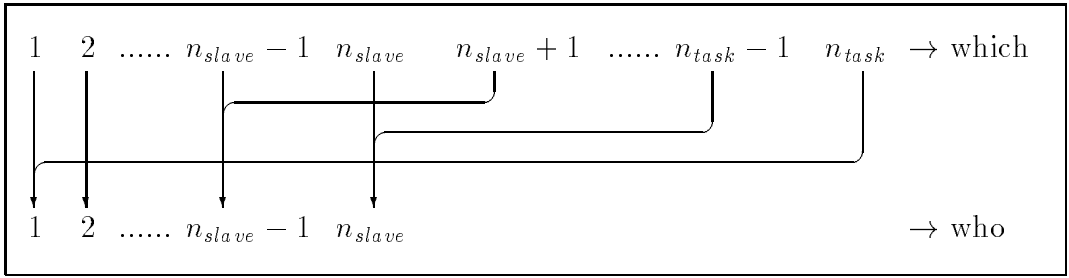

FIG. 7. Distribution of the processes between the slaves in the case of a distribution "at request".

```
procedure LOPOL3_MASTER (in data, n_atom)
begin
    CORRESPOND (n_atom, n_task, pratom, prdir)
    CONFIG (n_host)
    /* Determine the number of available processors */
    n_slave=MIN(n_task, n_host)
    /* Determine the number of slave */
    SPAWN (slave, n_slave)
    SEND (slave(1 → n_slave), {n_slave, slave, data, pratom, prdir})
    /* Send data and various informations to all slaves*/
    which=0
    do i_slave = 1 → n_slave
    /* Loop on the slaves */
        which=which+1
        SEND (slave(i_slave), {which})
        /* Send a first process to each slave */
    od
    do i_task = 1 → n_task
        RECEIVE (slave(1 → n_slave), {who})
        /* Wait until a slave has finished */
        if (which < n_task) then
            which=which+1
            SEND (slave(who), {which})
            /* Send a new process(which) to slave(who) */
        else
            SEND (slave(who), {who})
            /* Send an ending message to slave(who) */
        fi
    od
end
```

FIG. 8. Algorithm the LOPOL3 routine in the case of a distribution "at request".

The algorithms of LOPOL3 procedure (master) and of the program SLAVE_OF_LOPOL3 are reproduced in Figs. 8 and 9, using the generic language described in Fig. 4 for master/slaves communication.

```
program LOPOL3_SLAVE
begin
    RECEIVE (master, {n_slave, slave, data, pratom, prdir})
    /* Receive data and various informations from master*/
    ME (slave, n_slave, me)
    /* Determine who I am */
    RECEIVE (master, {which})
    /* Receive first process to treat */
    repeat
        i_atom=pratom(which)
        i_dir=prdir(which)
        /* Determine which perturbation I must treat */
        TREAT(data,i_atom,i_dir)
        SEND (master, {me})
        /* Send a message to the master to tell him I have finished */
        RECEIVE (master, {which})
        /* Wait for a new process or an ending message from master */
    until (which=me)
end
```

FIG. 9. Algorithm the program SLAVE_OF_LOPOL3 in the case of a distribution "at request".

Clearly, this method leads to a better load balancing on the processors. Moreover, it takes the pre-existing load of the processors into account. Indeed, the fastest processors will be the first to request a new task, while the slowest processors will not be overloaded.

Regarding the communications, the data are sent once to all processors at the beginning. Thus, the duration of $T_{comm}$ of the communications is independent of the number of processes $n_{task}$. This is quite interesting when this number increases.

Finally, let us discuss briefly how time measurement can be achieved in order to calculate the speedup and the efficiency. To be valid a time measurement (time elapsed between two given events) must be carried on a unique clock. Thus, to measure the time of communication $T_{comm}$ between two processes, synchronisation must be achieved. This can be done by two different means. Firstly, the sender starts a clock when he begins the communication, and stops it when he receives a signal from the addressee to tell him he got the message all right. Secondly, the addressee starts a clock when he receives a signal from the sender to tell him the communication is about to begin, and stops it when he got the message all right. In both cases, a signal must be sent; this means a new communication the duration of which should be known to have a precise time measurement. From a practical point of view, it will reveal useful only for long messages compared to the signal:

$$T_{message} \gg T_{signal}. \tag{1}$$

In general, the time measurement will be imprecise. For LOPOL3 procedure, the second method is easier to adapt to the "master & slaves" structure. Indeed, it avoids multiple clocks control by the master. The condition expressed by Eq. 1 suggests that precise time measurement can only be achieved for "Send data and various informations to all slaves". This is what was done using as a starting-communication signal one of the informations of the original message.

## II. THEORETICAL ANALYSIS

Let us know recall some elementary definitions and introduce notations. Let $T^*$ be the duration of the execution of the sequential code on a single processor, and $T(P)$ be the duration of the execution of the parallel code on $P$ processors. Then, the speedup $S(P)$ of the parallel code on $P$ processors is defined by:

$$S(P) = \frac{T^*}{T(P)}. \tag{2}$$

The efficiency $E(P)$ of the parallel code on $P$ processors is written as follows:

$$E(P) = \frac{S(P)}{P}. \tag{3}$$

Let us write $\alpha$ the fraction of the parallel code still executed sequentially on a single processor. We also define $T_{comm}(P)$ as the duration of the communications when the parallel code is executed on $P$ processors.

We now develop a theoretical model to evaluate the speedup of the parallel code taking into account the two sources of inefficiency that are communications and the remaining sequential fraction (Amdhal's law). Using the definitions above, we define $T_{calc}(P)$ by:

$$T(P) = T_{calc}(P) + T_{comm}(P), \tag{4}$$

it is the duration of numerical operations (excluding the communication) when the parallel code is executed on $P$ processors. It is clear that:

$$T_{calc}(1) = T^*, \tag{5}$$

while $T_{calc}(P)$ can be written as follows:

$$T_{calc}(P) = \alpha T_{calc}(1) + (1-\alpha)\frac{T_{calc}(1)}{P} \tag{6}$$

where we have supposed that an ideal speedup $P$ is obtained on the fraction $(1-\alpha)$ of the code executed in parallel $P$ processors. Finally, the speedup $S(P)$ is:

$$S(P) \leq P\left(\frac{1}{1+(P-1)\alpha}\right)\frac{1}{\left(1+\frac{T_{comm}(P)}{T_{calc}(P)}\right)}. \tag{7}$$

### III. APPLICATION TO SILICON

#### A. General Informations

The algorithms presented in this paper were implemented under PVM (version 3.2.2) [3–5]. The development and the preliminary tests were done on a cluster composed of one IBM RS6000-550 and one RS6000-590. The final timings were executed on a Convex MetaSeries consisting of 8 processors HP 735, each of which having a local memory of 128 MBytes. We decided to treat 8 perturbations in bulk silicon and analyze the speedup and the efficiency on 1,2,4 and 8 processors.

Our calculations are performed using a variational approach to density-functional perturbation theory [6], within the local density approximation (LDA) [7]. We use a rational polynomial parametrization of the exchange-correlation energy functional [8], which is based on the Ceperley-Alder gas data [9]. The iterative minimization procedure relies on a preconditioned conjugate gradient algorithm [10,11]. The electronic wavefunctions are sampled on a mesh of 10 special $k$ points in the IBZ and expanded in terms of a set of plane-waves whose kinetic energy is limited to 10 Hartree. The "all-electron" potentials are replaced by an *ab initio*, separable, norm-conserving pseudopotential built following the scheme proposed in Ref. [12].

#### B. Results

In the first test, we limite the number of iterations (only 2 minimization steps) in TREAT routine. This leads to a sequential time of $T^*$ =3450 s and to a quite high remaining sequential fraction $\alpha$ =5%. The resulting time measurements, speedup and efficiency are given in Table I. The drastic speedup reduction can be attributed to the sequential fraction $\alpha$, as shown by the second test hereafter. At the contrary, the loss of efficiency due to communications is almost negligible.

| $P$ | $T(P)$ | $T_{comm}(P)$ | $T_{calc}(P)$ | $S(P)$ | $E(P)$ |
|---|---|---|---|---|---|
| 1 | 3481 | 31 | 3450 | 0.99 | 99 % |
| 2 | 1837 | 31 | 1806 | 1.87 | 94 % |
| 4 | 1027 | 31 | 999 | 3.36 | 84 % |
| 8 | 620 | 31 | 589 | 5.56 | 70 % |

TABLE I. Results obtained with 1,2,4 and 8 processors in the case of bulk silicon (first test: limited number of iterations). Timings are expressed in seconds.

In the second test, we iterate in TREAT routine until convergence is reached (about 15 minimization steps), which is the real situation. This leads to a sequential time of $T^* = 24166$ s and to a nearly negligible remaining sequential fraction $\alpha = 0.7\%$. The resulting time measurements, speedup and efficiency are given in Table II. The measured speedup is plotted in Fig. 10 and is compared to the ideal speedup $S(P) = P$ and to the potential speedup that would be obtained if the sequential fraction was suppressed. The difference between the measured speedup, the potential one, and the ideal one shows that the loss of efficiency due to the sequential fraction $\alpha$ is negligible as well as that due to communications.

| $P$ | $T(P)$ | $T_{comm}(P)$ | $T_{calc}(P)$ | $S(P)$ | $E(P)$ |
|---|---|---|---|---|---|
| 1 | 24197 | 31 | 24166 | 0.99 | 99 % |
| 2 | 12166 | 31 | 12135 | 1.98 | 99 % |
| 4 | 6462 | 31 | 6431 | 3.74 | 94 % |
| 8 | 3356 | 31 | 3325 | 7.18 | 90 % |

TABLE II. Results obtained with 1,2,4 and 8 processors in the case of bulk silicon (second test: iterations until convergence). Timings are expressed in seconds.

It should be mentioned that the efficiency reduction that appears when the test is executed on 4 or 8 processors is due to the fact that 2 processes are shorter than the other. Thus 2 processors stay inactive waiting that the others have finished. It should also be noted that in both case the measured speedup is in good agreement with our theoretical model (see Eq. 7).

## IV. FINE-GRAIN PARALLELISATION

Each response computation is rather similar to one computation of the ground-state energy of the crystal, using the preconditioned conjugate gradient algorithm of Ref. [10,11]. The parallelisation of such an algorithm has already been studied by different groups [13–15]. The most fine-grained parallelism can be achieved at the level of the application of the

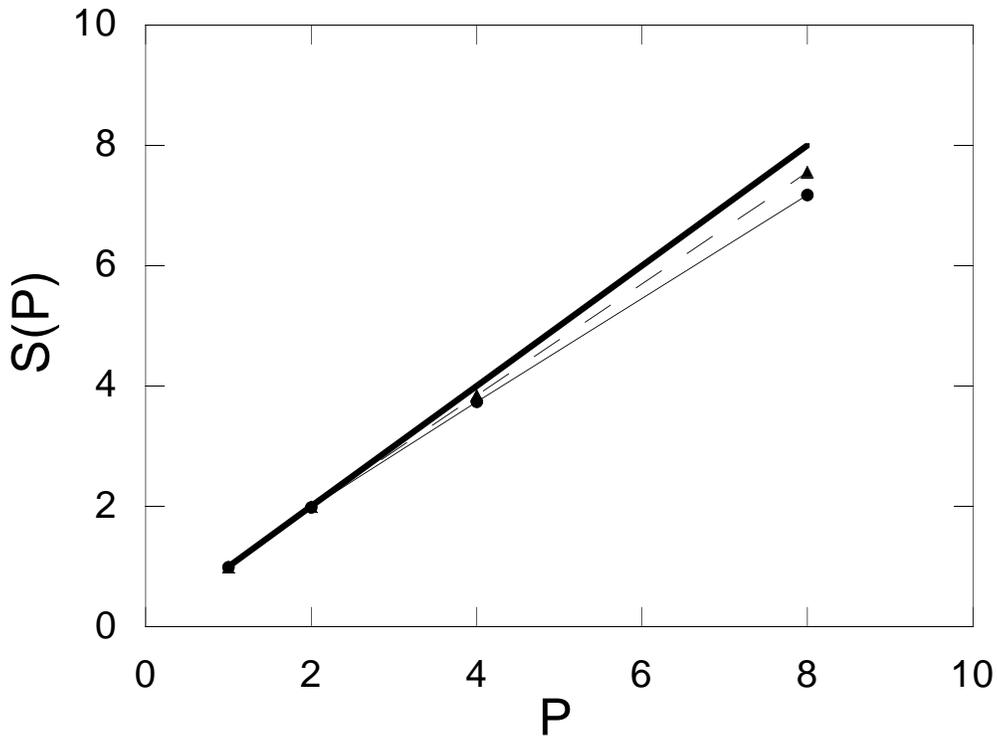

FIG. 10. Speedup obtained with 1,2,4 and 8 processors in the case of bulk silicon (second test: iterations until convergence). The solid line shows the limit of the ideal speedup $S(P) = P$. The circles represent the measured speedup, whereas the triangles are the potential speedup that would be obtained if the sequential fraction was suppressed (but still including the communication time).

Hamiltonian to the wavefunctions (similar to the multiplication of a sparse matrix and a vector). The number of coefficients describing the wavefunctions is usually between a few thousand and many millions, so that one has ample room for parallelisation along this line. Other parallelisations could be achieved by splitting the sequential computation of different k-points wavefunctions between different processors, or splitting the sequential computation of different bands between different processors, but we do not investigate them here.

In the Hamiltonian-to-wavefunction application, the most challenging part, as revealed in Ref. [13], is a three-dimensional Fast Fourier Transform, in which a large quantity data must be exchanged between different processors, for a comparatively modest amount of computation. We now describe the implementation of three-dimensional Fast Fourier Transform and its parallelisation.

Complex, double-precision, numbers are spread over a $N_x * N_y * N_z$ grid. A three-dimensional FFT involves Fourier transforming successively these data along the first direction, then along the second direction, then along the third direction: in the first step, $N_y * N_z$ independent one-dimensional (line) Fourier transform of size $N_x$ are performed; in the second step, $N_x * N_z$ independent one-dimensional (line) Fourier transform of size $N_y$ are performed; in the third step, $N_x * N_y$ independent one-dimensional (line) Fourier transform of size $N_z$ are performed. The grid is usually cubic, of size $24*24*24$ to $120*120*120$, typically. The one-dimensional Fast Fourier Transforms of size $N_x$,$N_y$, or $N_y$ are performed each on one processor using standard efficient sequential library routines. The parallelisation of the three-dimensional FFT algorithm involves distributing these one-dimensional FFT's accross the different processors, and making the output of each direction transform available to the processors in charge of the next direction transform.

In the Convex Exemplar architecture, the memory of one hyper-node is logically and physically shared between 8 processors. Each processor has a 1MB directly-mapped data cache attached to it. While the transfer of data between the cache and the processor is rather fast (1 cycle for a read), the communication between the memory and the cache is much slower (about fifty cycles for reading a 32-byte line cache). Thus, a central role is played by the management of the cache (data reuse), and the transfer of data between the memory and the cache, or, in the parallel case, between the different caches.

Let us first analyse the sequential algorithm, with respect to data transfer. If all the data can fit in the cache, then it can be loaded from the memory to the cache once, transformed along the three directions without further memory-to-cache transfer, then fetched back to the memory. As soon as the size of the grid becomes too large ($32*32*32$ complex, double-precision, numbers use already 512 kB, half the cache size), some additional transfer between the cache and the memory occurs. In order to minimize this transfer, one can load successively planes of data perpendicular to the z direction, and perform x and y direction Fourier Transforms within these planes before going to the next plane, then complete the z direction Fourier Transform once all data have been transformed in the x and y direction. Using this strategy, at most one supplementary complete transfer (write and read) of data between memory and cache occurs. Since the cache is large enough to contain all the data of a plane, further change of strategy is not required for the cases of interest (a plane of $512*512$ complex, double-precision, numbers use 1MB).

The parallelisation of the three-dimensional Fast Fourier Transform on the Convex Exemplar cannot maintain the low level of data transfer between memory and cache of the sequential algorithm, observed for the smaller grids. Indeed, suppose that two processors are available. The transforms along the x and y directions, for the lowest-half planes, are performed on processor number 1, while the transforms for the highest-half planes are performed on processor number 2. For this purpose, half of the data is transferred from the memory to the cache of processor 1, while the other half of the data is transferred from the memory to the cache of processor 2. After the x and y transforms are performed, the z transform cannot be performed unless some data is exchanged between the cache of processor 1 and 2. In this particular case, half of the data must be transferred between the processors at this intermediate stage. After the z transform has been done, the data must be written onto memory.

By contrast, parallelisation of three-dimensional FFT's for the larger grids involves no loss due to increased data traffic, because there was already a supplementary data transfer step.

To summarize: in trying to parallelize a three-dimensional Fast Fourier Transform on the Convex Exemplar,

- the number of numerical operations will be shared amongst different processors, thus allowing for potential speedup

- for the smallest grids, a supplementary transfer of data between memory and cache occurs

- for the larger grids, no such effect occurs, while the unavoidable transfers between memory and cache will be taken by different channels of communication (one for each processor), allowing also for speedup.

The implementation of this algorithm, and subsequent analysis of the timing confirms this analysis. The parallelisation of the code was easily obtained through the use of the C\$DIR LOOP_PARALLEL directive applied to the specific loop corresponding to the above mentioned algorithm, and compilation of the routine in which this directive appeared using the -O3 compilation option. The theoretical time to perform a three dimensional FFT of data spread on a cubic $N*N*N$ grid is proportional to $N^3 log_2 N^3$, when only floating point operations are taken into account. Since the logarithmic function varies very slowly, this time, divided by the number of data should stay nearly constant. On Fig. 11, we plot the mean time $T$ needed for one FFT, divided by $N^3$, as a function of the linear size of the grid, $N$, for $N = 16, 20, 24, 36, 40, 48, 60, 72$, using 1, 2 or 4 processors. With 1 processor, this quantity stays constant until $N = 40$ is reached. Then, it increases gradually. This is due to the cache effect previously discussed. By contrast, the 2-processor values are rather constant as a function of $N$. As a consequence, a small speedup (at most 1.4) is observed for values of $N$ smaller than 40, while it reaches 1.93 for $N = 60$ and 1.89 for $N = 72$. The 4-processor values are also rather independent of $N$, and the additional speedup with respect to the 2-processor value turns around 1.6.

From this analysis, we deduce that the fine-grain parallelisation based on a three dimensional FFT on a Convex Exemplar is more subtle than the coarse-grain parallelisation of response computations previously described. The memory to cache data transfer time can significantly affect the performance of the algorithm. As a consequence, the best sequential efficiency is obtained for the case of an amount of data that fits in the cache. In this case, the parallelisation is relatively inefficient. By contrast, when there are too much data to fit in the cache, the parallelisation speedup can be significant.

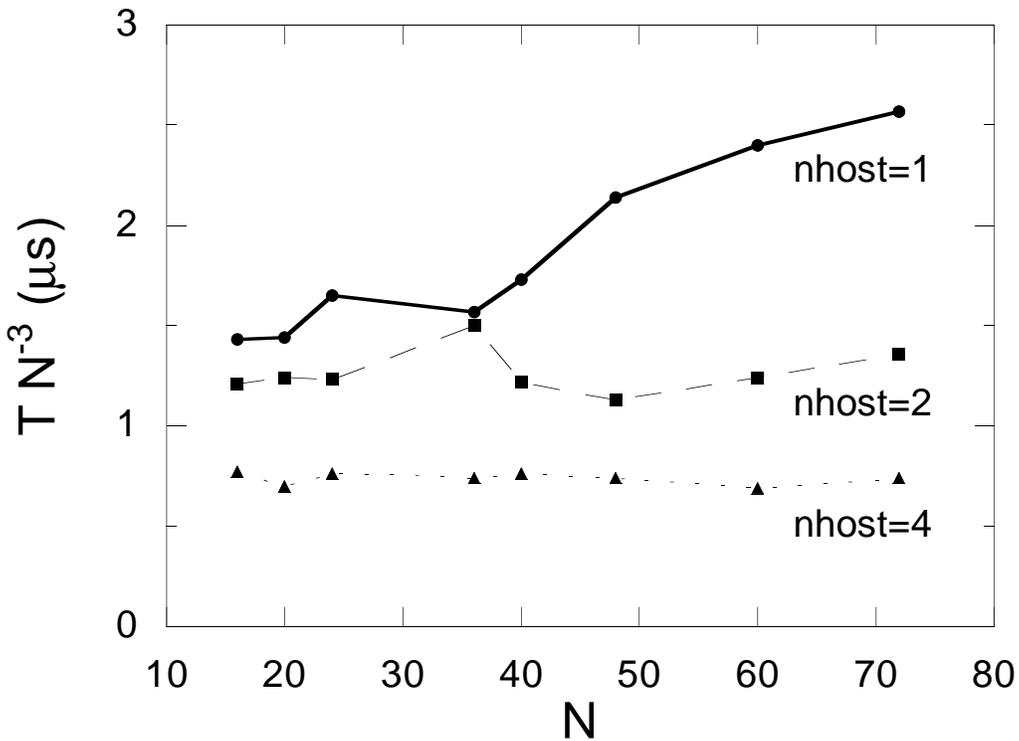

FIG. 11. Mean time $T$ needed for one FFT (expressed in $\mu$s), divided by $N^3$, as a function of the linear size of the grid, $N$, for $N = 16, 20, 24, 36, 40, 48, 60, 72$, using 1 (circles), 2 (squares) or 4 (triangles) processors.

Because of this mixed result, it is likely better to implement another, medium-grain, parallelisation, or to switch to another implementation of the basic equations of the response calculation [16].

## CONCLUSIONS

In this paper, we have first presented a coarse-grain parallel algorithm, based on "master & slaves" structure, for *ab initio* calculation of interatomic force constants. This parallel algorithm has lead to an excellent efficiency on a Convex MetaSeries computer, with up to 8 processors. This is due to the very low sequential fraction of the code and the relatively low amount of communications. Nevertheless, this parallelisation is too coarse for a large number of processors.

This is the reason why we have explored the parallelisation of the code, at a much finer level based on distributing a three-dimensional Fast Fourier Transform [FFT]. We have briefly described the parallelisation algorithm, its implementation on a Convex Exemplar. From the timing tests performed with up to 4 dedicated processors, it appeared that the cache behaviour is crucial. An interesting speedup was obtained for the larger size FFT's, although not as good as for the coarse-grain parallelisation.

For future development, an interesting approach would be to combine both coarse-grain and fine-grain parallelisations in order to get rid of the disadvantages of each of them.

## ACKNOWLEDGMENTS

The ground state results were obtained using a version of the software program Plane_Wave (written by D. C. Allan), which is marketed by Biosym Technologies of San Diego. Two of the authors (G.-M. R., and X. G.) have benefited from financial support of the National Fund for Scientific Research (Belgium). The Convex MetaSeries and Exemplar computer were made available thanks to a "Grand Projet Universitaire" of the UCL, and a "Action de Recherch Concertee" sponsored by the Belgian F.N.R.S. This paper presents research results of Belgian Program on Interuniversity Attraction Poles initiated by the Belgian State, Prime Minister's Office, Science Policy Programming.